%% file: paper.tex
\documentstyle[epsf]{EuroPhys}
\input EuroMacr

\begin{document}


\euro{}{}{-}{}
\Date{}
\shorttitle{S. ULREICH AND W. ZWERGER: 
GAUGE FIELDS IN A QUANTUM POINT CONTACT}

\title{Gauge Fields, Mode Mixing and Local Density of States 
in a Quantum Point Contact}
\author{S. Ulreich and W. Zwerger}
\institute{Sektion Physik, Ludwig-Maximilians-Universit\"at M\"unchen,\\ 
Theresienstra{\ss}e 37, D-80333 M\"unchen, Germany}

\rec{}{}

\pacs{
\Pacs{03}{80$+$r}{General theory of scattering}
\Pacs{72}{10$-$d}{Theory of electronic transport; scattering mechanisms}
}

\maketitle


\begin{abstract}
It is shown that the elimination of the  discrete transverse motion in
a    waveguide of arbitrary   shape may  be described  in   terms of a
non-abelian gauge field for the longitudinal dynamics. This allows for
an   exact  treatment of  the  scattering between   different modes by
eliminating the gauge field at the expense of a non-diagonal matrix of
local subband energies.  The method is  applied to calculate the local
density of states (LDOS) in a quantum point  contact.  Contrary to the
total   conductance which is    well     described by an     adiabatic
approximation, mode mixing turns out to play a  crucial role for local
properties like the LDOS.
\end{abstract}


The discovery of quantized longitudinal  conductances in quantum point
contacts  created    in  a  two-dimensional   ballistic  electron  gas
\cite{vanwees,wharam}  has   launched a great    number of theoretical
papers  dealing with quantum mechanical  scattering in waveguides with
a general  wide-narrow-wide type  geometry \cite{vanHouten}.   Precise
numerical calculations of the transmission amplitudes $t_{mn}$ between
the transverse  modes $m$  and  $n$  in  such a  structure  have shown
\cite{maao} that the adiabatic approximation $t_{mn} \sim \delta_{mn}$
-- while not exact -- works  extremely well for the linear conductance
\begin{equation}
\label{landauer}
  G = \frac{2e^2}{h} \sum_{nm} |t_{nm}|^2
\end{equation}
of  the point contact  at zero temperature.   The effect of scattering
between  different transverse  modes is therefore   negligible for the
conductance, thus justifying the adiabatic  approximations used in the
early  explanations \cite{glazman,yacoby} of the observed quantization
of $G$. In our present  work, we develop a  novel method dealing  with
quantum mechanical scattering in waveguides  of arbitrary width. It is
based  on  a gauge theoretic  description of  the effect of transverse
modes on the dynamics of the  longitudinal motion, which allows for an
exact treatment of intersubband  scattering.  Using this formalism and
the  recursive  Green function method  \cite{maao},  we  calculate the
local density  of states in a quantum  point contact.  In the vicinity
of the constriction  this local property is  strongly  affected by the
scattering between different transverse modes.

We consider a  general wave guide in two  dimensions with an arbitrary
but fixed   confining   potential $V(x,y)$.  To determine   the  exact
scattering wave functions of the stationary Schr\"odinger equation
\begin{equation}
  \left[ 
    -\frac{\hbar^2}{2 M} \nabla^2 + V(x,y)  
  \right] \psi(x,y) = E \psi(x,y)
\end{equation}
we first  split off the  discrete motion in the  transverse coordinate
$y$.   Quite  generally, for  any   confining potential $V(x,y)$ which
approaches infinity  as    $y \to  \pm  \infty$,  the  one-dimensional
Schr{\"o}dinger equation for the transverse motion  has a complete and
discrete basis  of eigenstates  $\Phi_n^x(y)$  for every value  of the
longitudinal coordinate $x$.   These eigenstates and eigenenergies are
of  course  position dependent.  The  position dependent eigenenergies
$\varepsilon_n(x)$  give rise to  potential barriers for the electrons
traversing the  wave guide structure in  longitudinal  direction.  The
corresponding eigenfunctions can  be chosen  real globally,  which  is
always possible in the absence  of a magnetic field.  The completeness
of the transverse  basis allows one to expand  the total wave function
in the form $\psi(x,y) = \sum_n \chi_n(x) \Phi_n^x(y).$ Inserting this
ansatz into the two-dimensional Schr\"odinger equation, we obtain
\begin{equation}
\label{withb}
\left[ \frac{\hat{p}^2}{2M} - \frac{i \hbar}{M} {\cal A}(x) \hat{p}
-\frac{\hbar^2}{2M} {\cal B}(x) + \underline{\underline{\varepsilon}}(x)
\right]
{\vec \chi}(x) = E {\vec \chi}(x).
\end{equation}
Here ${\vec \chi} (x) =  (\chi_1(x), \chi_2(x), \ldots)$ is the vector
of   the longitudinal   wave functions for    the different transverse
eigenstates $n$ and  $\hat p =  -i \hbar  \partial_x$  is the momentum
operator  for the longitudinal  motion.   To simplify the notation  we
have introduced the following matrices
\begin{equation}
\label{gaugedefinition}
{\cal A}_{nm} (x):= \int \limits_{-\infty}^{+\infty} \! dy \
\Phi_n^x(y) \cdot \frac{\partial \Phi_m^x(y)}{\partial x}
\quad
\mbox{and}
\quad
{\cal B}_{nm} (x):= \int \limits_{-\infty}^{+\infty} \! dy \
\Phi_n^x(y) \cdot \frac{\partial^2 \Phi_m^x(y)}{\partial x^2}
\end{equation}
and also the diagonal matrix of the local subband energies
\begin{equation}
{\varepsilon}_{nm} (x) := \langle n(x) | 
- \frac{\hbar^2}{2M} \frac{\partial^2}{\partial y^2} + V(x,y) | m(x) \rangle
 = \varepsilon_n(x) \cdot \delta_{nm} . 
\end{equation}
It is straightforward to see that $\cal A$ is antisymmetric, since the
wave  functions are real  valued everywhere.   Obviously ${\cal A}(x)$
and ${\cal B}(x)$   are  smooth for  confining  potentials,  which are
smooth  functions of  the  longitudinal coordinate $x$.   If there are
regions  in the form of  an ideal lead, we   have ${\cal A}(x) = {\cal
B}(x) = 0$ and the above system decouples trivially.

A  simple calculation  leads  to  the  following relation between  the
matrices ${\cal A}(x)$ and ${\cal B}(x)$
\begin{equation}
{\cal B}(x) = \frac{d {\cal A}(x)}{dx} + {\cal A}^2(x),
\end{equation}
which  was first  discovered by Kuperin  {\it  et al.} \cite{kuperin}.
This relation allows one to rewrite eq. (\ref{withb}) in the form
\begin{equation}
\label{hamiltoneq}
\left[ 
\frac{1}{2 M} \left( \hat{p} - i \hbar {\cal A} \left(x\right) \right)^2
+ \underline{\underline{\varepsilon}}(x)
\right] {\vec \chi}(x) = E {\vec \chi}(x).
\end{equation}
Formally   this   is  an   tensorial  Hamiltonoperator  acting   on an
infinite-dimensional vector of wave functions. Apparently, the form of
our Hamiltonian now resembles  the quantum mechanical Hamiltonian of a
charged particle in a magnetic field, however, in the present case the
gauge field $\cal A$ is a tensor instead of a vector.

We  now  consider unitary   transformations performed   on the set  of
transverse eigenfunctions,  i.e.      we  introduce a    local   gauge
transformation (the bar denotes complex conjugation)
\begin{equation}
\label{local_gauge_wave}
| \tilde{k}(x) \rangle = \sum_n \overline{S}_{kn}(x) | n(x) \rangle,
\end{equation}
which mixes the different transverse eigenmodes.  Since we do not want
to change the  physics via this operation,  we also have to change our
longitudinal coefficients according to
\begin{equation}
\tilde{\chi}_k(x) = \sum_l S_{kl}(x) \chi_l(x),
\end{equation}
leaving the entire wave function invariant.   It is straightforward to
show,   how   the differential   equation  (\ref{hamiltoneq}) for  the
longitudinal   coefficients   ${\vec  \chi}(x)$   changes  under  this
transformation.  The transformation  law  for the diagonal  matrix  of
subband energies $\underline{\underline{\varepsilon}}(x)$ is simply
\begin{equation}
\label{energy_transform}
\underline{\underline{\tilde{\varepsilon}}}(x)=
S(x) \underline{\underline{\varepsilon}}(x) S^+(x).
\end{equation}
The new and   old gauge potentials    $\tilde{\cal A}(x)$ and   ${\cal
A}(x)$, however, are   related by the  more complicated transformation
law
\begin{equation}
\label{potential_transform}
\tilde{\cal A}(x) = S(x) {\cal A}(x) S^+(x) - \frac{dS(x)}{dx} S^+(x),
\end{equation}
well-known  from  non-abelian  gauge  theories  \cite{itzykson}.   The
transformed Schr\"odinger-equation   for the longitudinal coefficients
then has the same form as (\ref{hamiltoneq}), i. e.
\begin{equation}
\label{transformeddeq}
\left[ 
\frac{1}{2 M} \left( \hat{p} - i \hbar \tilde{{\cal A}}(x) \right)^2
+ \tilde{\underline{\underline{\varepsilon}}}(x)
\right] \tilde{{\vec \chi}}(x) = E \tilde{{\vec \chi}}(x).
\end{equation}
Since $\psi(x,y)$ is   unchanged, all observables are invariant  under
this  gauge transformation. Choosing a  gauge in this context means to
choose a  certain basis  of  transverse wave  functions at every space
point $x$.

The formalism above can now be used to describe the mixing between the
different transverse modes -- which is formally described by the gauge
potential ${\cal A}(x)$  -- in a very  explicit form, that  also turns
out to be convenient  for a numerical  treatment.  To this end a gauge
transformation is performed to a gauge  in which $\tilde{{\cal A}}(x)$
vanishes identically.  From (\ref{potential_transform}) this  requires
that $S(x){\cal A}(x) = dS(x)/dx$ or
\begin{equation}
\frac{dS^+(x)}{dx}S(x) = - {\cal A}(x).
\end{equation}
This differential equation for the transformation matrix $S(x)$ can be
solved formally by
\begin{equation}
\label{formal}
S(x,x_0) = R \left[ 
\exp \int \limits_{x_0}^x \! d \xi \ {\cal A}(\xi) 
\right],
\end{equation}
where $R$ is the space-ordering  operator, defined in complete analogy
to the  time-ordering operator known  from time-dependent perturbation
theory.  The antisymmetry of the gauge  potential $\cal A$ ensures the
unitarity of the evolution operator $S(x,x_0)$. In order to understand
the physics behind our special gauge, we note that $\cal A$ is defined
by
\begin{equation}
\frac{\partial}{\partial x} | n(x) \rangle = 
- \sum_m {\cal A}_{nm}(x)| m(x) \rangle .
\end{equation}
The transverse basis  in the gauge  where $\tilde{{\cal A}}(x) = 0$ is
therefore obviously the one with a fixed  reference point $x_0$.  Thus
by expanding the total wavefunction in terms of transverse modes which
are  independent   of  the longitudinal   coordinate    $x$, the gauge
potential clearly vanishes.  The  fixed basis set $\Phi_n^{x_0}(y)$ is
related to  the locally adiabatic  basis $\Phi_n^{x}(y)$ at  any point
$x$   by  the  unitary     transformation  $S(x,x_0)$ as     given  in
(\ref{formal}).     As   a  result,   the   original   diagonal matrix
$\underline{\underline{\varepsilon}}(x)$  of  the  adiabatic   subband
energies is transformed to a  non-diagonal form via the transformation
law (\ref{energy_transform}).  Mode-mixing is thus no longer described
by  a non-vanishing    gauge potential but    is  instead due   to the
off-diagonal    terms      in   the     transformed     energy  matrix
$\tilde{\underline{\underline{\varepsilon}}}(x)$.

For an application  of the general  theory  above, it is  necessary to
specify the confining potential. In order to obtain analytical results
as far as possible, we choose a harmonic confinement of the form
\begin{equation}
V(x,y) = \frac{M}{2} \omega_0^2 \left( \frac{b_0}{b(x)} \right)^2 y^2.
\end{equation}
The effective width $b(x)$ has a minimum $b_0 =  b(x=0)$ at the origin
and widens  to a finite value $b_\infty  =  b(x\to\pm\infty)$ at $|x|$
larger than a characteristic constriction length $L$. Quite generally,
for a harmonic confining  potential, the gauge potential ${\cal A}(x)$
may be written in the form
\begin{equation}
{\cal A}(x) = \frac{1}{4} \frac{b'(x)}{b(x)} 
\left ( \hat{a}^{+ \, 2} - \hat{a}^2 \right)
\end{equation}
where $\hat{a}$ is the usual  harmonic oscillator destruction operator
and the  prime  denotes differentiation with respect  to   $x$.  Since
${\cal  A}(x)$ factorizes into a  simple function  of  $x$ and a fixed
operator, the spatial ordering symbol  in (\ref{formal}) is irrelevant
in the present case. The  evolution operator $S$ is therefore obtained
in explicit form as
\begin{equation}
S(x,x_0) = \exp \left[ 
\frac{1}{4} \ln \left( \frac{b(x)}{b(x_0)} \right )
\left( \hat{a}^{+ \, 2} - \hat{a}^2 \right)
\right].
\end{equation}
This   operator  turns  out to   be   identical  with   the  so called
squeeze-operator well known in   quantum optics \cite{squeeze}.    Its
matrix  elements in  the  transverse  oscillator  eigenstates  can  be
expressed  analytically   \cite{phd}     in terms  of   hypergeometric
functions, which are not given here explicitly, however.

In order   to  solve the    coupled  differential equations   for  the
longitudinal wave functions  ${\vec \chi}(x)$, we  eliminate the gauge
potential and  discretize  the transformed equation  in  steps of size
$a$, such that ${\vec  \chi}(x=ia) = {\vec \chi}_i$. Transforming back
to    the  original   adiabatic    basis with     a  diagonal   matrix
$\underline{\underline{\varepsilon}}(x)$  of  subband   energies,  the
resulting equations are
\begin{equation}
\label{tight_bind}
\left(\underline{\underline{\varepsilon}}_{\; i} + 2 \right){\vec \chi}_i
-S_{i+1,i}{\vec \chi}_{i+1}  - S_{i-1,i}{\vec \chi}_{i-1}
= E {\vec \chi}_i  
\end{equation}
in an   obvious notation  for $\underline{\underline{\varepsilon}}_{\;
i}$,    $S_{i+1,i}$ and   with  dimensionless energies     in units of
$\hbar^2/(2Ma^2)$. This is a tight-binding like model in one dimension
with   as many orbitals  per site  $i$   as number of transverse modes
included  (about 20 in  the numerical calculations).  A convenient and
numerically  stable method to solve it  is the recursive Greenfunction
technique \cite{maao}. This allows one  to calculate the matrix of the
local Greenfunctions $G_{mn}(i,j)$ at  a given  energy $\varepsilon_F$
from which the local density of states
\begin{equation}
\rho(x,y; \varepsilon_F) = -\frac{1}{\pi}
\sum_{mn} \Phi^x_m(y) \Phi^x_n(y)  \cdot  \mbox{Im} \; G_{mn}(x,x)
\end{equation}
is finally obtained.  This  quantity  can be  directly measured in  an
experiment with  a scanning tunneling  microscope (STM),  provided the
two-dimensional electron gas can be realized on a free surface like in
InAs-systems.   Indeed the local tunneling current  between an STM tip
and  a  conducting sample is directly   proportional  to $\rho(\vec x;
\varepsilon_F)$ \cite{STM}. In   the standard case where the  electron
gas resides below  a surface layer,  the LDOS may  be determined via a
local capacitance spectroscopy as was recently demonstrated by Ashoori
 \cite{ashoori}.    An example for  the local   density of
states is shown in fig.~1 for a smooth boundary function $b(x)$ with a
width    ratio $b_0/b_\infty=0.1$ as   indicated in   the figure.  The
characteristic  length  $L$ is  $250  \ nm$  while  the  Fermi  energy
$\varepsilon_F  = \hbar \omega_0/2$ is choosen  such that it coincides
with  the lowest  transverse   mode  at  the   narrowest  point  $x  =
0$. Including the quantum mechanical tunneling and reflection over the
barrier \cite{glazman}, the  associated linear conductance is equal to
$e^2/h$, i.e. we are at the inflection point of the transition between
pinch  off and the  first conductance plateau  at $G=  2 e^2/h$.  This
situation is nicely reflected  in the  exact  calculation of  the LDOS
shown in fig.~1 on the left, where a narrow dip at the constriction is
just  about  to be closed.  In  the  wide regime where four transverse
modes  are occupied, the  LDOS exhibits a complex interference pattern
reflecting the  square  of the   exact wavefunction $\sum_n  \chi_n(x)
\Phi_n^x(y)$   at  the Fermi energy.  In   this  asymptotic regime the
structure is reproduced  rather well   in an adiabatic   approximation
where all matrices $S$ in   (\ref{tight_bind}) are replaced by a  unit
matrix.  The resulting LDOS  is shown in fig.~1  on  the right for the
same parameter values.  It is very similar to the  exact result in the
regime where $b(x)$ is  constant.  Near the constriction, however, the
adiabatic  approximation  obviously  differs  strongly  from the exact
result and in particular it  fails to  reproduce the almost  pinch-off
structure.  Mode mixing   therefore plays a   crucial role for   local
properties like the LDOS.

\begin{figure}[t]
\centering
\noindent
\begin{minipage}[t]{.48\linewidth}
\centering   
\leavevmode
\epsfxsize = \linewidth
\epsffile{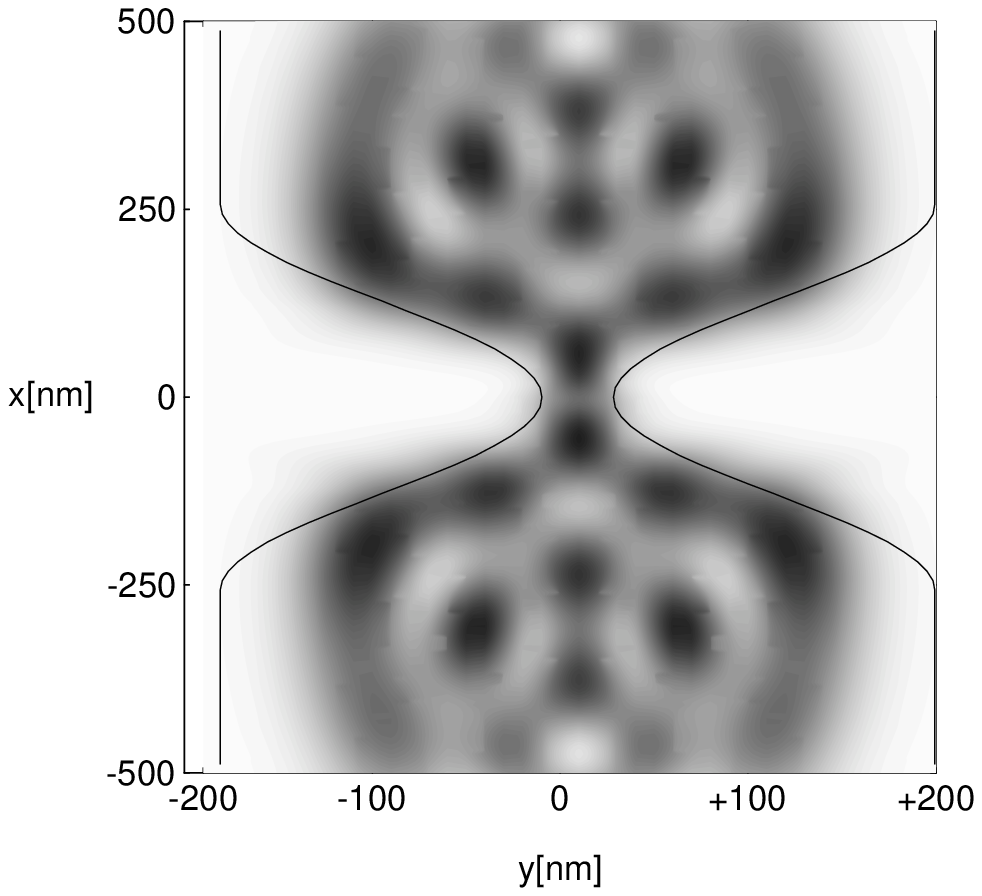}
\end{minipage}
\begin{minipage}[t]{.48\linewidth}
\centering  
\leavevmode
\epsfxsize = \linewidth 
\epsffile{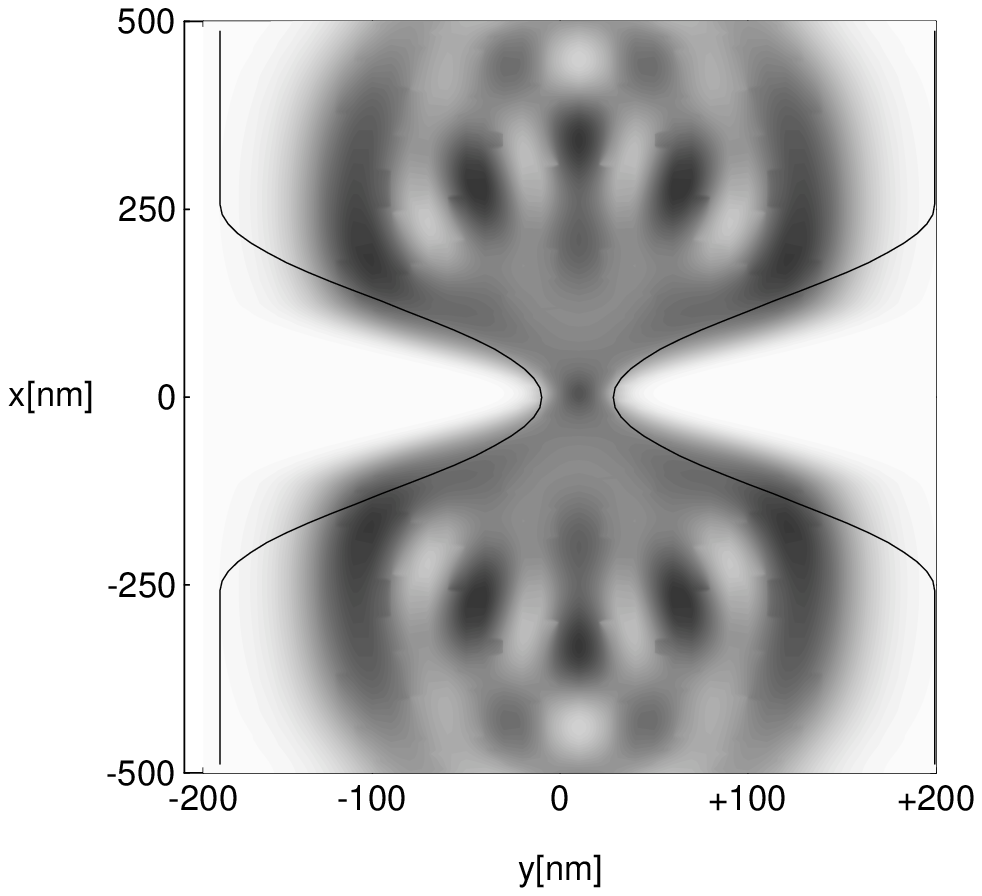}
\end{minipage}
\caption{Local density of  states for a  quantum point contact  with a
harmonic confinement of width $b(x)$ as  indicated. In the vicinity of
the constriction  the  ``exact'' calculation (left)  differs  strongly
from the result obtained in an adiabatic approximation (right).}
\end{figure}

In  conclusion we have  developed   a novel  method  to treat  quantum
scattering in general wave guides via  a gauge field approach and have
applied it for a calculation of  the LDOS in  a quantum point contact.
The nontrivial local  structure found here   may possibly be  observed
using modern  scanning probe techniques.  Our method  can also be used
to determine  the local potential  distribution in  the presence of  a
finite  current  \cite{ulreich}  and   the   local   emissivities  and
injectivities  which  determine   the
ac-transport properties at low frequencies \cite{buettiker_wechsel}.


\stars    
It  is a pleasure  to acknowledge useful  discussions with Bert Lorenz
and David Wharam about the  possibilities of measuring  the LDOS in  a
quantum point contact.   This work was  supported by the  {\sc SFB 348
Nanometer-Halbleiter-Bauelemente}.\\
\stars


\end{document}

%% file: EuroMacr.tex
\def\And{{\rm and\ }}

\def\stars{\bigskip\centerline{***}\medskip}

\newif\ifboo \boofalse

\def\Review#1{\boofalse{\it #1},}
\def\Name#1{{\sc #1},}
\def\Vol#1{\ifboo Vol. {\bf #1}\else{\bf #1}\fi}
\def\Year#1{\ifboo #1\else(#1)\fi}
\def\Book#1{\bootrue{\it #1},}
\def\Page#1{\ifboo {\rm p. #1}\else{\rm #1}\fi}